# Manifestation of Lepton Interaction Nonuniversality in Spontaneously Violated Mirror Symmetry

*Igor T. Dyatlov \**
*Scientific Research Center "Kurchatov Institute"*
*Petersburg Institute of Nuclear Physics, Gatchina, Russia*

LHC$_b$ data indicate a significant difference in widths between the $B \to K$ (or $K^*$) $+ e^+e^-$ and $B \to K$ (or $K^*$) $+ \mu^+\mu^-$ decays (April 2017). The width of the $e^+e^-$-channel is noticeably larger than that of the $\mu^+\mu^-$-channel. SM-calculations require equality here. The difference may mean that a new interaction exists which changes generations and distinguishes leptons, with coupling constants much larger than, and inverse in power to, the SM-coupling of fermions with the Higgs boson. In the spontaneously violated mirror symmetry, the coupling between SM particles and the heavy (in principle) second Higgs scalar, necessarily present here, is characterized by precisely such properties. The inverse strong power of such a coupling is not an additional hypothesis but also a necessary condition for the qualitative reproduction of the observed mass hierarchy of charged leptons and the structure of lepton weak mixing, the *PMNS* matrix. In the mirror model being discussed, all properties, including the inverse power of the new interaction, are dictated by the hierarchical character of mass spectra for quarks and charged leptons.

## 1. Introduction

Recently published more precise LHC$_b$ data for *B*-meson decays [1] indicate that a new source of lepton interaction nonuniversality other than the coupling with the Higgs boson *H* has possibly been discovered. We reason that at this time, ratios between the widths of the *μ*- and *e*- channels, where uncertainties of strong processes are reduced, reflect most truthfully the properties of this possible new physics. Results and interpretation of direct width measurements and angular distributions [2] do not appear to be final. The partial widths of the $B \to K$ (or $K^*$) $+ e^+e^-$ and $B \to K$ (or $K^*$) $+ \mu^+\mu^-$ decays are noticeably different from each other, while their Standard Model (SM) computations should be almost identical despite mass differences [3]. Of note is the larger width of decays into electrons versus *μ*-muons. This last property may be interpreted as the constants of the new interaction should be inverse to the constants of the standard coupling with *H*. The influence of *H* on the processes discussed is negligible.

This discovery sparked considerable interest; dozens of interpretations of the phenomenon followed instantly (arXiv, April-May 2017).

In this paper, we are offering one more interpretation. In [4-8], in order to explain the structure of weak mixing matrices (WMM) for quarks (the *CKM* matrix) and leptons (the *PMNS* matrix) [9], a model based on the assumption that SM particles have heavy mirror analogs was proposed.

---

*E-mail: dyatlov@thd.pnpi.spb.ru

Mass hierarchies of quarks and charged leptons appear to be an observed SM characteristic which is crucial for reproduction of all qualitative properties of both quarks and leptons [4, 8]. The weak isospin *SU*(2) symmetry also plays an important role.

Mirror symmetry (MS) violation may be achieved in a similar fashion to SM [10]—that is, using isodoublet scalars forming vacuum condensates. However, MS per se is a theory where the system must be able to choose spontaneously either the left-handed (*L*) or right-handed (*R*) weak currents of light particles. This property can be achieved only by introducing two scalars [6]. The first is the normal Higgs isodoublet that produces masses of weak vector bosons and the Higgs boson *H*. The second presents four new particles (0,-,+,0) which may have very heavy masses.

This new boson $\Phi_2$ has the ability to participate in strong interactions with SM particles, with power inverse to their masses. This is not an additional condition imposed on the system but rather a necessary consequence of MS and the observed phenomenology (mass hierarchy). The interaction of leptons with this particle is precisely what may lead to lepton nonuniversality.

A possible confirmation of this hypothesis may be furnished by the *B* (and $B_s$) → $e^+e^-$ or $\mu^+\mu^-$ decays [9] with SM contributions according to the chiral rules being minor. These processes may be expected to have similar properties, i.e., more pronounced $\Phi_2$ contributions to the *e*-channel. At present, the experiment is on the brink of discovering these decays (the $\mu^+\mu^-$ decay is considered already discovered [11] although with low confidence).

In Section 2, the model [4-8] is used to complete a transition from MS fermions of the initial Lagrangian to particles that have masses. In Section 3, we investigate the coupling of SM mass states with the new isodoublet $\Phi_2$ and prove that the power of this interaction is inverse to SM's Yukawa constants that govern the coupling with the Higgs boson *H*. Appendix 1 describes the appearance of SM's *H*-constants in the MS scenario, which is here nontrivial. Appendix 2 discusses parity nonconservation in MS models.

## 2. Transition from mirror symmetrical states to states with masses

In the MS scenario presented in [4-8], the fermion, quark and lepton states of the system's Lagrangian are expressed only in terms of the isodoublets $\Psi_{LR}$ and isosinglets $\Psi_{RL}$:

$$\Psi_{LR} = \psi_L + \Psi_R \left(T = \frac{1}{2}\right), \quad \Psi_{RL} = \psi_R + \Psi_L \left(T = 0\right). \tag{1}$$

The chiral parts (*R, L* – right, left) of these massive fermions are written in a form that provides visual representation of mirror symmetry. Eq.(1) assumes three generations and two flavors $f = \bar{u}, \bar{d}$.

Besides the mass terms (4), this Lagrangian is the sum of two SMs for $\psi$ and $\Psi$ which differ from each other as a result of $R \leftrightarrow L$ substitution. Violation of the MS

$$\psi \leftrightarrow \Psi \quad L \leftrightarrow R \tag{2}$$

can be achieved by introduction of Yukawa couplings with two isodoublet scalars $\Phi_1$ and $\Phi_2$ in a similar fashion to SM [6]. One of the scalars produces the condensate $\langle \Phi \rangle = \eta \simeq 246$ GeV, providing only $\Phi$ and $\psi$ masses. The system can exist in one of the two states which must be identical in all properties except weak ones and, partially, Yukawa couplings where $L$ universally changes to $R$. Identity is an inherent attribute of MS required to prevent fixation of the right- or left-handed system by physical means without an independent determination of the $\Psi$- and $\psi$-difference.

As a result of MS (2) violation, the initial Lagrangian for the quark or charged lepton system[1] consists, omitting the SM terms, of two parts [7].

1. The terms defined by the Yukawa couplings and by the formation of $\Psi$ fermion mass matrices:

$$h_a^b \bar{\Psi}_R^a \Psi_{Lb} \eta + h_a^b \bar{\Psi}_R^a \Psi_{Lb} H + h_a^b (\bar{\psi}_L \Phi_2) \psi_{Rb} + c.c. , \tag{3}$$

$H$ is the Higgs boson of SM, $h$ is the matrix of the Yukawa coupling—generally speaking, an arbitrary matrix of the generation indices $a, b = 1, 2, 3$. Similar to SM, we have for one of the flavors ($f = \bar{d}$): $\Phi_2 \rightarrow \Phi_2^C = i\sigma_y \Phi_2^*$. The matrix $h$ is the same for all terms (3). It is the condition for the identity of both possible states in an MS system after MS violation.

The Yukawa couplings (3) are presented in a non-diagonal form similar to SM, although this nondiagonality, for the chosen model, could be directly transferred to the objects in item 2, below.

2. Masses of the states (point 1) connecting $\psi$ and $\Psi$ components:

$$\tilde{A}_a \bar{\Psi}_{LR}^a \Psi_{LRa} + \tilde{B}_b \bar{\Psi}_{RL}^b \Psi_{RLb} = \tilde{A}_a \bar{\psi}_L^a \Psi_{Ra} + \tilde{B}_b \bar{\psi}_R^b \Psi_{Lb} + c.c. \tag{4}$$

By choosing nondiagonal $h$, the parameters $\tilde{A}$ and $\tilde{B}$ can be considered as diagonal real matrices. Weak *SU(2)* invariance requires that the isodoublet masses $\tilde{A}$ be independent of the flavor $f$: $\tilde{A}^{(\bar{u})} = \tilde{A}^{(\bar{d})}$, while the isoscalar masses $\tilde{B}$ may depend on $f$: $\tilde{B}^{(\bar{u})} \neq \tilde{B}^{(\bar{d})}$.

The transition from the components $\Psi$, $\psi$ to the massive particles $\chi$ and $\chi_{SM}$ can be completed in two steps.

1. Diagonalize the Yukawa coupling matrices $h^{(f)}$, and consequently the mass matrices of $\Psi$ particles: $\mu = h\eta$. In the MS model, this can be done even prior to MS violation by transformation of all operators (1) and conversion of diagonal $\tilde{A}$ and $\tilde{B}$ into nondiagonal matrices $A$ and $B$.

---

[1] A somewhat more complicated procedure for neutrino [5] is analogous.

This operation does not result here in a WMM for SM particles, since the nondiagonal isodoublet matrix $A$ must be independent of flavor, as were the $\tilde{A}$ masses. The reason for this is the weak $SU$(2) symmetry and phenomenology; only in this case, the quark WMM structure (the hierarchy of *CKM* matrix elements) can be reproduced in full and without any additional prerequisites. The WMM is here a result of a different mechanism [4,5].

Further, the Yukawa constants $h$ and the unitary matrices diagonalizing them (in a general case two for each $f$) cannot be arbitrary. The most general form for $h$ is written out as follows:

$$\Psi_{LR} = U_{LR}\Psi_{LR}^{(d)}, \quad \Psi_{RL} = U_{RL}\Psi_{RL}^{(d)},$$
$$h^{(f)} = U_{LR}h_{(diag)}^{(f)}U_{RL}^{(f)+}. \tag{5}$$

where $\Psi_{LR(RL)}^{(d)}$ are the eigenfunctions of $h$. The unitary isodoublet matrix $U_{LR}$ does not depend on $f$. At that, the weak Lagrangian remains diagonal, and this is the principal difference from the classical SM procedure.

The matrices $A$ and $B$ are equal

$$A = U_{LR}^{+}\tilde{A}U_{LR} \quad B^{(f)} = U_{RL}^{(f)+}\tilde{B}^{(f)}U_{RL}^{(f)}, \tag{6}$$

they are Hermitian, and $A$ is independent of $f$. $U$ transformations do not affect the diagonal forms of all other interactions in SM.

The operation (5) sets the problem: to express (3)-(4) in terms of massive states to obtain the matrix of the 6th order

$$M_{LR} = \begin{vmatrix} \mu & B^+ \\ A & 0 \end{vmatrix}, \tag{7}$$

where $\mu$ is a diagonal matrix, and $A$ and $B$ are Hermitian matrices (3 x 3). To reproduce the properties of the SM WMM for quarks and charged leptons, all $\mu_n$ elements have to be much larger than $A$ and $B$ elements, i.e., the eigenvalues of these matrices, $\bar{A}$ and $\bar{B}$ [4]:

$$|\mu| \gg |\tilde{A}|, |\tilde{B}|. \tag{8}$$

The matrix (7) is a direct generalization of the see-saw mechanism [12]. Besides the three large masses $\mu_n$, it also has three small eigenvalues [13]. To single them out, let us find a matrix inverse to (7). We obtain precisely:

$$M_{LR}^{-1} = \begin{vmatrix} 0 & A^{-1} \\ B^{+-1} & -B^{+-1}\mu A^{-1} \end{vmatrix}. \qquad (9)$$

2. From (9) we can see that, in the zero approximation, the mass matrix for small masses is:

$$m_a^b = -\sum_{n=0}^{2} A_a^n \frac{1}{\mu_n} B_n^{+b}. \qquad (10)$$

Diagonalizing the separable matrix (10) is the second step in the problem of transition to SM massive particles. In Eq.(10), $a, b = 1,2,3$ are indices of $\psi$ generations in the space of diagonal $\mu$. The numbers $n = 0,1,2$ for generations of mirror particles (as in [4,5]) were chosen in order to solve the problem of diagonalization (10) by expanding by the parameters of the observed SM mass hierarchy (quarks $f = \bar{u}, \bar{d}$ and charged leptons). Let us assume that the SM mass hierarchy is defined by the inverse mass hierarchy of mirror particles $\mu_n$, while the difference between the parameters of the matrices $A$ and $B$ for various generations is less essential. Then, inequality (8), given the form of the mass matrix (10), can be rewritten in a more illustrative way:

$$\frac{|\tilde{A}| \text{ or } |\tilde{B}|}{\mu} \simeq \left(\frac{|AB|}{\mu^2}\right)^{1/2} \simeq \left(\frac{m_{SM}}{\mu}\right)^{1/2} \ll 1, \qquad (11)$$

where $m_{SM}$ is masses of SM particles. The lightest mass $\mu_0$ corresponds in (10) to the zero approximation for the observed hierarchy, i.e., it defines in (10) the heaviest SM particles and so on:

$$\mu_2 \gg \mu_1 \gg \mu_0. \qquad (12)$$

In [4], we found the masses and eigenvalues of matrix (10) using the perturbation theory (12). It was shown that condition (12) and the independence of the matrix $A$ from flavor define in (10) a mass spectrum of SM particles with a hierarchy inverse to (12):

$$m_I \ll m_{II} \ll m_{III}, \qquad (13)$$

and a quark WMM (the *CKM* matrix) with the Wolfenstein hierarchical pattern [14]. In (13), the numbering $I = I, II, III$ represents indices of ordinary SM generations.

With conditions (8) and (11) fulfilled, Eq.(1) corresponds to the following distribution of mirror particle masses and SM states:

$$m_{III} \ll \mu_0, \qquad (14)$$

that is, all mirror states are significantly heavier than SM states. This is one of the conditions for reproduction of WMM properties [9], which distinguishes this MS model from other attempts to introduce mirror systems (see [15]).

The role of subsequent terms of expansion by minor parameters (8), (11) is discussed in Appendix 1. For our discussion here, it is sufficient to consider the lowest orders of expansion.

## 3. Interaction Distinguishing Leptons

The new interaction distinguishing $e, \mu, \tau$ is the coupling of $\psi$-components with the second isodoublet $\Phi_2$ (3). The interaction with the Higgs boson *H*, which also distinguishes leptons, is not significant for light particles of SM and is not considered in our scenario (see Appendix 1).

As discussed in [6], the mass $\Phi_2$ may be indefinitely high, and therefore the quantitative evaluation of nonuniversality through this coupling is not possible. In addition, the large (generally speaking, nonperturbative) value of the Yukawa coupling with mirror states in the scenario being discussed also hinders such evaluation. This coupling is responsible here simultaneously for huge masses of mirror particles, their coupling with the Higgs boson (Appendix 1) and the coupling of SM states with the boson $\Phi_2$, since the necessary condition of MS is the equality of Yukawa constants for $\Phi_1$ and $\Phi_2$; consequently:

$$h \simeq \frac{\mu}{\eta} \gg 1, \tag{15}$$

At the same time, the large $\Phi_2$ mass requires a strong coupling with a noticeable effect, as it may be observed in the *B*-decay processes. Under these conditions, we can only establish the relative values of the $\Phi_2$ coupling with various leptons (or quarks) of SM. The inverse mass hierarchy of mirror particles must support increased $\Phi_2$ interaction with the lightest particles of SM, which makes this coupling attractive for explanation of the results [1].

Using the unitary matrices (5) diagonalizing $h^{(f)}$ and the matrices $V_L$ and $V_R$ (obtained in [4]) diagonalizing (10), we can write out $\psi_L$ and $\psi_R$ in Eq.(3) in terms of states with defined masses $\chi_L^{(SM)}$ and $\chi_R^{(SM)}$:

$$\psi_L = U_{LR} V_L \chi_L^{(SM)}, \quad \psi_R = U_{RL} V_R \chi_R^{(SM)}. \tag{16}$$

In [4], the matrices $V_L$ diagonalizing (10) were found only for the left states $\chi_L^{(SM)}$, however, the right matrices $V_R$ of the separable matrix (10) can apparently be expressed through the same formulas used for the left matrices, with $A \leftrightarrow B$ substitution.

Substituting (15) into (3), we obtain for the coupling with $\Phi_2$:

$$h(\bar{\psi}_L \Phi_2)\psi_R = (\bar{\chi}_L^{(SM)} \Phi_2)^I V_{LI}^{+n} (U_{LR}^+ h U_{RL})_n^{n'} V_{Rn'}^{I'} \chi_{RI'}^{(SM)}$$
$$\equiv (\bar{\chi}_L^{(SM)} \Phi_2)^I V_{LI}^{+n} \frac{\mu_n}{\eta} V_{Rn}^{I'} \chi_{RI'}^{(SM)}, \quad (17)$$

where $U_{LR}^+ h U_{RL}$ is a diagonal mass matrix $\mu_n$ of mirror particles; $I$ and $I'$ are, again, indices of observed lepton generations; and $n = 0,1,2$ are indices of mirror lepton generations.

For the mass hierarchy $\mu_n$ (Eq.(12)) inverse to the SM mass spectrum, the greatest contribution to (17) comes from the term with the maximally large $\mu_n \equiv \mu_2$. At that, the rows of the matrices $V_L$ and $V_R$ corresponding to the largest mass $n = 2$ contain a large matrix element $V_2^1 \simeq 1$ responsible for the transition to SM's lightest particles, i.e., electrons. Other elements of this row (as well as column) of the unitary matrix are small, being in the order of $(|\tilde{A}|$ or $|\tilde{B}|)/\mu \ll 1$. Such structure $V$ results entirely from the hierarchical properties of spectra and expansions by hierarchy powers.

Let us prove this statement.

The first stage of diagonalization $h$ can occur if we neglect all elements except the one that defines the largest mass (12). Assuming that there is only one such element (similar to the see-saw mechanism), let us place it in the upper left-handed corner of the matrix $h$ (by changing the numeration of the generation indices $a, b$). This very form $h$ leads in [8] to the observed structure of WMM leptons—that is, the *PMNS* matrix. We will show below that to achieve this result, the simplified Hermitian form $h$ used in [8] is not required. Of importance is only the hierarchical structure of the spectrum.

The eigenfunctions of $h$ (for both the $L$ and $R$ components) for the heaviest $\mu_n \equiv \mu_2$ (12) will apparently have the following form:

$$\phi_{L,R} = \frac{1}{N} \begin{pmatrix} \sim 1 \\ 0_1 \left( \frac{|\tilde{A}| \text{ or } |\tilde{B}|}{\mu} \right) \\ 0_2 \left( \frac{|\tilde{A}| \text{ or } |\tilde{B}|}{\mu} \right) \end{pmatrix} \simeq \begin{pmatrix} 1 \\ \sim 0 \\ \sim 0 \end{pmatrix}. \quad (18)$$

Such functions are included in the matrices $U_{LR}$ and $U_{RL}$. The largest mass $\mu_2$ in matrix (10) corresponds to the $A_2$ and $B_2$ vectors in the generation space of diagonal to $\mu_n$. In turn, $A_2$ and $B_2$, calculated using Eq.(6) with matrices containing column (18), will be "almost" orthogonal to the vectors $A_0$ and $A_1$ or, respectively, $B_0$ and $B_1$. In the lowest approximation of the hierarchy, the eigenfunctions of the separable matrix (10) depend only on the $A$ vectors for $L$ states or only on the $B$ vectors for $R$ states [4]. The matrix elements $V_L$ and $V_R$ are scalar products of these $L$ and $R$ functions. The normalized vectors in the lowest order of the hierarchy (13)

$$\frac{1}{N}\left[A_0^*, A_1^*\right] \text{ and } \frac{1}{N}\left[B_0^*, B_1^*\right], \tag{19}$$

([…] being the vector product) are correct eigenfunctions for the zero approximation of the matrix (10) for the $L$- and $R$-states with the smallest mass $m_I$ [4]. They are "almost" coincident with the "directions" of $A_2$ and $B_2$ whose unit vectors are $L, R$-directions of the heaviest mirror states of the heaviest mirror masses $\mu_2$. Consequently, we have

$$V_{L2}^I(A), \ V_{R2}^I(B) \simeq 1. \tag{20}$$

Thus, the interaction (17) is the strongest for couplings with electrons ($u$- and $d$-quarks). Other fermions of SM have much weaker couplings with $\Phi_2$—either through the lower masses $\mu_0, \mu_1$, or through $\mu_2$ but small elements of the $V_{L,R2}$ vectors:

$$\sim \left(|\tilde{A}| \quad \text{or} \quad |\tilde{B}|\right)/\mu \sim (m_{SM}/\mu)^{1/2}.$$

As mentioned above, it is impossible to evaluate quantitatively the influence of (17) on the real processes due to possible nonperturbativity and the unknown mass of the scalar $\Phi_2$. It is clear, however, that such interaction would result in differences of cross-sections with participation of various leptons, exerting a greater influence on electrons than on $\mu$- and $\tau$-leptons. The processes $B \to K(K^*) + e^+ e^-$ and $B \to K(K^*) + \mu^+ \mu^-$ are possible examples of such influence. If this is true, the $K\mu^+\mu^-$ process with a very small contribution from (17) should coincide with SM. In magnitude, it is the second order of weak interaction. In the analogous $Ke^+e^-$ decay, a part of the width would be dependent on $\Phi_2$.

A study of the $B(B_S) \to e^+ e^-$ decays, in which the $\Phi_2$ mechanism should be more pronounced than that in $B(B_S) \to \mu^+\mu^-$ decays, could confirm the suggested role of the MS-$\Phi_2$ coupling. The latter decays have already been observed, albeit with low confidence:

$$Br(B \to \mu^+\mu^-) = (3.9^{+1.6}_{-1.4}) \cdot 10^{-10} \, [9], \ Br(B_S \to \mu^+\mu^-) = (3.0 \pm 0.6^{+0.3}_{-0.3}) \cdot 10^{-9}. \, [11]$$

It is asserted that these values are in agreement with SM estimates. For $B(B_S) \to e^+e^-$, limits much greater than the $\mu^+\mu^-$ width are only available:

$$Br(B \to e^+e^-) < 8.3 \cdot 10^{-9} \, [9], \ Br(B_S \to e^+e^-) < 2.8 \cdot 10^{-7} \, [9].$$

In the reaction $B_S \to e^+e^-$ ($b \to S$ transfer, as in [1]), the presence of $\phi_2$ could become obvious starting from $Br \lesssim 10^8$. In the analogous decays of the $B^0$, $K^0$, $D^0$ hadrons, contributions can be provided by other parts of the interaction (17).

The $\Phi_2$ mechanism could also result in processes involving flavor change such as $e\bar{\nu}$, $\mu\bar{\nu}$, etc. In this case, however, the major, much greater contribution to their cross-sections is provided by the first orders of weak interaction. Weakened $(\sim \sqrt{m_{SM}/\mu})$ processes with nonconservation of lepton numbers such as $\bar{e}\mu$ are also possible.

The mechanism under consideration (~ second order) cannot be responsible for the significant deviation of the $B \to D^*\tau\bar{\nu}$ decays (~ first weak order) from SM, see [16]. However, the most recent Belle results [17] demonstrate that the very existence of such deviations requires further confirmations.

## Appendix 1

Let us consider the role of the next in order terms of matrix (7) expansion by the parameter (8), (11). At arbitrary $\mu$, *A*, *B*, computation of even the next approximation is cumbersome and overcomplicated, which prevents a clear picture of consequences that ensue. The qualitative characteristics and structure of results are clearly demonstrated in an analytical solution where the eigenvalues $\mu_n$ are close to each other:

$$\left|\frac{\mu_n - \mu_{n'}}{\mu}\right| \sim \frac{|\tilde{A}|^2 \quad \text{or} \quad |\tilde{B}|^2}{\mu^2} \ll 1. \tag{A1}$$

Then, to calculate the matrix *M*, (7), we preserve the terms $(|\tilde{A}|$ or $|\tilde{B}|)/\mu$ and neglect the terms (A1). At Hermitian *A* and *B*, for the approximation being considered, the sixth-order matrices

$$X_A^+ = \begin{vmatrix} 1 & (1/\mu)A^+ \\ -A(1/\mu) & 1 \end{vmatrix}, \quad X_B = \begin{vmatrix} 1 & -(1/\mu)B^+ \\ B(1/\mu) & 1 \end{vmatrix} \tag{A2}$$

are unitary. They transform Eq.(9) to the following form:

$$X_A^+ M X_B = \begin{vmatrix} \mu + \frac{1}{\mu}A^+A + B^+B\frac{1}{\mu} & \frac{1}{\mu}A^+A\frac{1}{\mu}B \\ -A\frac{1}{\mu}B^+B\frac{1}{\mu} & -A\frac{1}{\mu}B^+ \end{vmatrix}. \tag{A3}$$

For the approximation being considered (A1), we obtain the block-diagonal matrix:

$$\mathcal{M} = \begin{vmatrix} \mu & 0 \\ 0 & -A\frac{1}{\mu}B^+ \end{vmatrix}. \tag{A4}$$

Complete diagonalization of all of the mass terms (3) and (4) is apparent here—this occurs with the already known matrices $U_{LR}, U_{RL}, V_L, V_R$ with supplement (A2). For initial fermions and operators $\Psi$ and $\psi$, couplings with the massive states $\chi$ and $\chi^{(SM)}$ are written out as follows:

$$\Psi_L = U_{RL}\left(\chi_L - \frac{1}{\mu} A^+ V_L \chi_L^{(SM)}\right), \quad \psi_L = U_{LR}\left(V_L \chi_L^{(SM)} + A\frac{1}{\mu}\chi_L\right);$$

$$\Psi_R = U_{LR}\left(\chi_R - \frac{1}{\mu} B^+ V_R \chi_R^{(SM)}\right), \quad \psi_R = U_{RL}\left(V_R \chi_R^{(SM)} + B\frac{1}{\mu}\chi_R\right),$$

(A5)

where $\chi$ are the states of heavy mirror particles. Here, similar to (7), $\mu$ is a diagonal mass matrix of mirror particles. The values $U_{RL}, V_R$, $B$ and $\mu$ depend on the flavor $f$. Eqs.(A5) permit determination of properties of all interactions in the MS model under consideration [4].

Interaction with the Higgs boson $H$ is of greatest interest. In the initial Lagrangian of the broken MS (3), $H$ interacts with the $\Psi$-components only. Substituting Eq.(A5) into (3) we obtain a diagonal coupling of $H$ with mass particles $\chi$ and $\chi^{(SM)}$ that exactly corresponds to SM:

$$\bar{\Psi}_L^n h_n^{n'} \Psi_{Rn'} H = \bar{\chi}_L^I U_{LRI}^{+n} h_n^{n'} U_{RLn'}^{I'} \chi_R H + \bar{\chi}_L^{I(SM)} V_{LI}^{+n} A_a^b \frac{1}{\mu_b} U_{LRb}^{+n} h_n^{n'} U_{RLn'}^{b'}$$

$$\times \frac{1}{\mu_{b'}} B_{b'}^{+\,a'} V_{Ra'}^{I'} \chi_{RI'}^{(SM)} H + \binom{nondiagonal}{part} = \chi_L^n \frac{\mu_n}{\eta} \chi_{Rn} H + \chi_L^{I(SM)} \frac{m_I}{\eta} \chi_{RI}^{(SM)} H + \binom{nondiagonal}{part}.$$

(A6)

The diagonality of the first term of the sum (A6) is obvious. In the second term of the sum, we obtain the matrix $A\frac{1}{\mu}B^+$, which is diagonalized precisely by the matrices *V*. For these contributions, there are no restrictions related to the terms neglected in (A1). In [6] we indicate that such interaction must be precise for the Higgs boson—this is a direct consequence of weak *SU*(2) symmetry violation. We can see from (A6) that interaction of mirror particles with the Higgs boson $H$ is, generally speaking, nonperturbative, since $\mu_n \gg \eta$. Therefore, formulae with participation of mirror particles $\chi$ have for most part illustrative purposes.

The nondiagonal part (A6) of the interaction with *H* in (3) describes mirror particle decays $\chi \to H + (SM)$:

$$-\bar{\chi}_R U_{LR}^+ h U_{RL} \frac{1}{\mu} A^+ V_L \chi_L^{(SM)} H - \bar{\chi}_R^{(SM)} V_R^+ B \frac{1}{\mu} U_{LR}^+ h U_{RL} \chi_L H \equiv$$

$$\equiv -\bar{\chi}_R \frac{A^+}{\eta} V_L \chi_L^{(SM)} H - \bar{\chi}_R^{(SM)} V_R^+ \frac{B}{\eta} \chi_L H.$$

(A7)

This coupling can be perturbative due to the fallout of very large mirror masses. The interaction of this very type can be the major reason for the instability of mirror particles. For the hierarchical spectrum of mirror leptons and quarks inverse to SM, the largest elements $V_L$ and $V_R$ are related with decays of the heaviest particles into the lightest states of SM fermions (Eq.(20), i.e., electrons or $u, d$ quarks.

For weak interactions we have, firstly, standard terms, defined by the principal terms in (A5) and as such independent of the character of the spectrum $\mu_n$ and restrictions (A1):

$$g\bar{\Psi}_R^a(\gamma\tau W)\Psi_{Ra} + g\bar{\psi}_L^a(\gamma\tau W)\psi_{La} \rightarrow g\bar{\chi}_R^n(\gamma\tau W)C_\mu\chi_{Rn'} + g\chi_L^{I(SM)}(\gamma\tau W)C_m\chi_{LI'}^{(SM)} + \ldots \quad \text{(A8)}$$

where $C_m$ is a mixing matrix; $C_m = 1$ for neutral quark or lepton currents. The mixing matrix of mirror particles is:

$$C_\mu \simeq \delta_{nn'} + 0\left(\frac{|\tilde{A}|^2 \text{ or } |\tilde{B}|^2}{\mu^2}\right), \quad C_\mu \equiv 1$$

for neutron currents.

Secondly, we have terms here that are responsible for weak decays of mirror particles:

$$-g\bar{\chi}_R(\gamma\tau W)\frac{1}{\mu}B^+V_R\chi_R^{(SM)} - g\chi_L^{(SM)}V_L^+(\gamma\tau W)A\frac{1}{\mu}\chi_L + c.c. \quad \text{(A9)}$$

According to (8), these contributions are much smaller than the weak ones.

In addition, Eq.(A8) leads to very small right-handed currents of SM particles and left-handed currents of heavy mirror fermions:

$$g\,\bar{\chi}_R^{(SM)}V_R^+B\frac{1}{\mu}(\gamma\tau W)\frac{1}{\mu}B^+V_R\chi_R^{(SM)} + g\bar{\chi}_L\frac{1}{\mu}A^+(\gamma\tau W)A\frac{1}{\mu}\chi_L. \quad \text{(A10)}$$

The interaction with the second isosinglet $\Phi_2$ from the Lagrangian (9) in the main part, which is of interest for modern physics, is discussed in the main body of the paper.

## Appendix 2. Parity Nonconservation in MS Model

The MS model is built such that, provided there are no differences between the fermion components $\Psi$ and $\psi$ in (1), the parity *P*, i.e., the symmetry $\vec{x} \rightarrow -\vec{x}$, $R \rightarrow L$, is conserved.

Lepton phenomenology [5] requires that the Majorana masses of neutrino correspond to parity conservation even after MS violation ($M_R = -M_L$).

It appears attractive and natural that parity nonconservation in the MS model should occur only through weak interactions and mass difference.

Upon MS breaking and with differences between $\Psi$ and $\psi$, the change $\Psi \leftrightarrow \psi$ means a change of the isospin states $T = 0 \leftrightarrow T = 1/2$. This appears to conserve *P* in all parts of the Lagrangian that are diagonal by flavor $f = \bar{u}, \bar{d}$: kinetic terms, diagonal gauge interactions. The interaction with the Higgs scalar conserves *P* for the Hermitian matrix $h$ of the Yukawa constants. However, the second isospinor, necessary in MS, violates the parity in the Yukawa coupling of the broken system even for the Hermitian $h$. Therefore, the assertion in [5] that the Hermitian $h$ represents *P* violation only in weak interactions upon MS breaking by scalar bosons is not correct.

This paper shows that $h$ being non-Hermitian does not affect the conclusions of [5] regarding the structure of the *PMNS* matrix arising out of the MS model. The property necessary for this— the approximate orthogonality of the vector $A_2$ related with the largest mass of mirror particles, to the vectors $A_0$ and $A_1$—results exclusively from the hierarchical character of mass spectrum and does not depend on the hermiticity of $h$.